\documentclass
[twocolumn,showpacs,showkeys,fleqn,10pt,secnumarabic,rmp,final]{revtex4}%
\usepackage{amsfonts}
\usepackage{amsmath}
\usepackage{amssymb}
\usepackage{graphicx}
\usepackage{times}%
\begin{document}
\title{Spin precession in inversion-asymmetric two-dimensional systems}
\author{Ming-Hao Liu}
\email{d92222010@ntu.edu.tw}
\affiliation{Department of Physics, National Taiwan University, Taipei 106, Taiwan}
\author{Ching-Ray Chang}
\email{crchang@phys.ntu.edu.tw}
\affiliation{Department of Physics, National Taiwan University, Taipei 106, Taiwan}

\begin{abstract}
\hrulefill

We present a theoretical method to calculate the expectation value of spin in
an inversion-asymmetric two-dimensional (2D) system with respect to an
arbitrarily spin-polarized electron state, injected via an ideal point
contact. The 2D system is confined in a [001]-grown quantum well, where both
the Rashba and the Dresselhaus spin-orbit couplings are taken into account.
The obtained analytical results allow more concrete description of the spatial
behaviors of the spin precession caused by individually the Rashba and the
Dresselhaus terms. Applying the calculation on the Datta-Das spin-FET, whose
original design considers only the Rashba effect inside the channel, we
investigate the possible influence due to the Dresselhaus spin-orbit coupling.
Concluded solution is the choice of $\pm$[1$\pm$10], in particular [110], as
the channel direction.

\hrulefill

\end{abstract}
\date{10 March, 2006}
\maketitle

\section{Introduction}

The Datta-Das spin field-effect transistor (spin-FET) \cite{1}, though not yet
experimentally realized, has stimulated plenty of research on related topics
in the emerging field of semiconductor spintronics \cite{2,3}. The original
design contains a two-dimensional electron gas (2DEG) channel, bridged by
ferromagnetic spin injection source and detection drain contacts. Considering
the structure inversion asymmetry (SIA) of the confinement potential, a
spin-orbit (SO) coupling due to the Rashba effect \cite{4} generates an
effective magnetic field, which is always perpendicular to the electron
propagation inside the channel. Through an applied gate voltage, the strength
of the confinement potential and the corresponding Rashba field may be varied,
thus tuning the spin precession rate for the injected electrons, transiting
along the 2DEG channel. The stronger the gate voltage applied, the higher the
spin precession rate is. Therefore, the spin orientation angle for the
electrons arriving at the end of the channel, and hence the resulting current,
is theoretically tunable via the applied gate voltage. Such proposed device is
expected to serve as a field-effect transistor based on the electron spin, and
hence called Datta-Das spin-FET.

The principal advantage of such a spin-FET is that it may switch faster than
the traditional one since the charge redistribution is no more involved.
However, realization of the Datta-Das spin-FET faces basic challenges,
including \cite{3}: (i) effective controllability of the Rashba SO coupling
strength $\alpha$, (ii) long spin-relaxation time in 2DEGs, (iii) uniformity
of $\alpha$, and (iv) more efficient spin injection rate. So far, the former
two conditions have been basically satisfied in experiments \cite{5,6}, while
the latter two remain to be solved. Furthermore, recent attention has even
been put on the possible influence due to the bulk inversion asymmetry (BIA)
\cite{7} of the underlying crystal containing the 2DEG, which is neglected in
the original design of Datta and Das \cite{8}.

Previously, a great part of attention has been put on the spin injection rate
for the ferromagnet-2DEG junction structures \cite{9,10}. To examine the
feasibility of the Datta-Das spin-FET theoretically and more generally,
however, we turn to the fundamental problem, namely, the spin precession due
to space inversion asymmetry in two-dimensional systems.

In this paper, we begin with a quantum mechanical approach of detection for an
arbitrarily polarized and perfectly injected electron spin inside a 2DEG
channel, with both the SIA and BIA involved. Applying such techniques we
analyze the spatial behaviors of the spin precession due to individually the
SIA, the BIA, and simultaneously the SIA and BIA. With proper orientation of
spin injection, we can thus examine the 2DEG channel of the Datta-Das spin-FET
and see what influence the additionally considered BIA may bring.

\section{Quantum-mechanical spin detection}

The logic of how the space inversion asymmetry can lead to spin precession is
quite simple and clear. Let us summarize as follows. From Kramers theorem, one
has degenerate energy bands $E_{\uparrow\left(  \downarrow\right)  }\left(
-\mathbf{k}\right)  =E_{\downarrow\left(  \uparrow\right)  }\left(
\mathbf{k}\right)  $ in solids when no external magnetic field is applied
\cite{11}. With space inversion symmetry, i.e., the energy bands are
degenerate along $\pm\mathbf{k}$ for a certain spin state $\uparrow$ or
$\downarrow$: $E_{\uparrow\left(  \downarrow\right)  }\left(  -\mathbf{k}%
\right)  =E_{\uparrow\left(  \downarrow\right)  }\left(  \mathbf{k}\right)  $,
one is led to the Kramers degeneracy $E_{\uparrow}\left(  \mathbf{k}\right)
=E_{\downarrow}\left(  \mathbf{k}\right)  $. In other words, a zero-field
spin-splitting $E_{\uparrow}\left(  \mathbf{k}\right)  \neq E_{\downarrow
}\left(  \mathbf{k}\right)  $ will exist intrinsically when the crystal does
not own the space inversion symmetry. The spin-splitting, though does not stem
from the applied magnetic field, can be ascribed to an effective magnetic
field, leading to spin precession about the field direction.

Put in another way, the structure inversion asymmetry of the crystalline
structure is the key to the spin precession, and hence the principle of the
Datta-Das spin-FET. As mentioned in the previous section, the SIA of the
confinement potential plays the crucial role in the Datta-Das spin-FET for
gate-voltage tunability and can be described in the linear Rashba model
\cite{4} by%
\begin{equation}
H_{R}=\dfrac{\alpha}{\hbar}\left(  p_{y}\sigma_{x}-p_{x}\sigma_{y}\right)
\text{,}%
\end{equation}
choosing the 2DEG growth direction [001] along $z$. Symbols $\sigma_{x}$ and
$\sigma_{y}$ are the well-known Pauli matrices. When the 2DEG is confined in a
zinc-blend-based semiconductor, the BIA (the Dresselhaus term \cite{7})
described by the linear term%
\begin{equation}
H_{D}=\dfrac{\alpha}{\hbar}\left(  p_{x}\sigma_{x}-p_{y}\sigma_{y}\right)
\text{,}%
\end{equation}
must be further considered. Parameter $\beta$ depicts the strength of the
Dresselhaus SO coupling and is material-specific. Note that the contribution
of the $k^{3}$ term in the Dresselhaus SO coupling is neglected since the 2DEG
plane is assumed to be thin. The full Hamiltonian describing the electron%
\begin{equation}
H=\dfrac{p^{2}}{2m^{\ast}}+H_{R}+H_{D}%
\end{equation}
in the effective mass approximation then leads to the eigenfunctions%
\begin{equation}
\left\langle \mathbf{r}|\mathbf{k},\sigma\right\rangle =\dfrac{e^{i\mathbf{k}%
\cdot\mathbf{r}}}{\sqrt{2}}\left(
\begin{array}
[c]{c}%
ie^{-i\varphi}\\
\sigma
\end{array}
\right)  \equiv e^{i\mathbf{k}\cdot\mathbf{r}}\left\vert \psi_{\sigma
}\right\rangle \text{,}%
\end{equation}
with%
\begin{equation}
\varphi\equiv\arg\left[  \left(  \alpha\cos\phi+\beta\sin\phi\right)
+i\left(  \alpha\sin\phi+\beta\cos\phi\right)  \right]  \text{,}%
\end{equation}
corresponding to the eigenenergies%
\begin{equation}
E_{\sigma}\left(  k\right)  =\dfrac{\hbar^{2}k^{2}}{2m^{\ast}}+\sigma\gamma
k\text{,}%
\end{equation}
where the composite SO\ strength $\gamma$ is defined by%
\begin{equation}
\gamma=\sqrt{\alpha^{2}+\beta^{2}+2\alpha\beta\sin\left(  2\phi\right)
}\text{.}%
\end{equation}
Note that the space part and the spin part in the eigenfunction Eq. (4) are
entangled, and can not be expressed as direct products. By this we mean that
the spinor always contains the information of the wave vector, i.e., the
electron propagation angle $\phi$.

Let us now inject a spin-polarized electron described by%
\begin{equation}
\left\vert s_{\text{inj}}\right\rangle _{\mathbf{r}_{0}}=\dfrac{1}{\sqrt{2}%
}\left(
\begin{array}
[c]{c}%
e^{-i\phi_{s}}\\
1
\end{array}
\right)  \text{,}%
\end{equation}
on $\mathbf{r}_{0}$ in 2DEG. Expanding Eq. (8) in terms of the
Rashba-Dresselhaus eigenstates Eq. (4), we have%
\begin{equation}
\left\vert s_{\text{inj}}\right\rangle _{\mathbf{r}_{0}}=\sum_{\sigma=\pm
1}c_{\sigma}e^{i\mathbf{k}_{\sigma}\cdot\mathbf{r}_{0}}\left\vert \psi
_{\sigma}\right\rangle \text{,}%
\end{equation}
where the expansion coefficients $c_{\sigma}$ are given by $\left\langle
\psi_{\sigma}|s_{\text{inj}}\right\rangle _{\mathbf{r}_{0}}$. When the
electron is spatially evolved to another point, say $\mathbf{r}$, the state
ket then reads%
\begin{equation}
\left\vert s_{\text{inj}}\right\rangle _{\mathbf{r}_{0}\rightarrow\mathbf{r}%
}=\sum_{\sigma=\pm1}c_{\sigma}e^{-i\mathbf{k}_{\sigma}\cdot\left(
\mathbf{r}-\mathbf{r}_{0}\right)  }\left\vert \psi_{\sigma}\right\rangle
\text{,}%
\end{equation}
which is the consequence of operating the translation operator $\exp\left(
-i\mathbf{p}\cdot\mathbf{r}/\hbar\right)  $. When factoring out a phase
$\exp\left[  i\left(  \mathbf{k}_{+}-\mathbf{k}_{-}\right)  \cdot\left(
\mathbf{r}-\mathbf{r}_{0}\right)  \right]  $, which does not affect any
physical quantity, we can express the state ket evolved from $\mathbf{r}_{0}$
to $\mathbf{r}$ as%
\begin{equation}
\left\vert s_{\text{inj}}\right\rangle _{\mathbf{r}_{0}\rightarrow\mathbf{r}%
}=\sum_{\sigma=\pm1}c_{\sigma}e^{-i\Delta\theta\left(  \mathbf{r}%
-\mathbf{r}_{0}\right)  /2}\left\vert \psi_{\sigma}\right\rangle \text{,}%
\end{equation}
where we have utilized the fact that the difference of the Fermi momenta along
a certain direction, say $\phi$, is a fixed value%
\begin{equation}
\left(  \mathbf{k}_{+}-\mathbf{k}_{-}\right)  \cdot\hat{r}_{\phi}%
=\dfrac{2m^{\ast}\gamma\left(  \phi\right)  }{\hbar^{2}}\text{,}%
\end{equation}
and the phase difference is defined by%
\begin{equation}
\Delta\theta\left(  \mathbf{r}\right)  =\dfrac{2m^{\ast}\gamma}{\hbar^{2}%
}r\text{.}%
\end{equation}
Using Eq. (11) and setting $\mathbf{r}_{0}=0$, the expectation values of spin
operators $S_{x}$, $S_{y}$, and $S_{z}$ can be obtained as%
\begin{equation}
\left[
\begin{array}
[c]{c}%
\left\langle S_{x}\right\rangle \\
\left\langle S_{y}\right\rangle \\
\left\langle S_{z}\right\rangle
\end{array}
\right]  =\dfrac{\hbar}{2}\left[
\begin{array}
[c]{c}%
\cos\phi_{s}\cos^{2}\dfrac{\Delta\theta}{2}-\cos\left(  2\varphi-\phi
_{s}\right)  \sin^{2}\dfrac{\Delta\theta}{2}\\
\sin\phi_{s}\cos^{2}\dfrac{\Delta\theta}{2}-\sin\left(  2\varphi-\phi
_{s}\right)  \sin^{2}\dfrac{\Delta\theta}{2}\\
\cos\left(  \varphi-\phi_{s}\right)  \sin\Delta\theta
\end{array}
\right]  \text{.}%
\end{equation}
The above spin vector expression is suitable for an inplane-polarized spin
[Eq. (8)] injected on $\mathbf{r}_{0}$, and is functions of detection position
$\mathbf{r}$, Rashba and Dresselhaus coupling strengths $\alpha$ and $\beta$,
and the orientation angle of spin polarization $\phi_{s}$. When considering
the general case with arbitrary spin polarization for the injected electrons,
one need only replace the spinor expressed in Eq. (8) by%
\begin{equation}
\left\vert s_{\text{inj}}\right\rangle _{\mathbf{r}_{0}}=\left(
\begin{array}
[c]{c}%
e^{-i\phi_{s}}\cos\left(  \theta_{s}/2\right) \\
\sin\left(  \theta_{s}/2\right)
\end{array}
\right)  \text{,}%
\end{equation}
where $\theta_{s}$ is the polar angle.

\section{Spatial behaviors}

We now demonstrate the spatial behaviors of the spin precession due to
different mechanisms of inversion asymmetry by using the analytical formulae
Eq. (14). Consider a 2DEG plane with spin injection on the center of the
plane. The polarization of the injected spin is set parallel to [100]. Assume
that the electron possess a conserved wave vector and is free to move in the
2DEG plane along any crystallographic direction.

Let us begin with the Rashba case, i.e., only the SIA mechanism of inversion
asymmetry is present. The well-known Rashba field is circularly polarized and
the corresponding spin precession is shown in Fig. 1(a). Such a circular
polarization of the induced effective magnetic field manifests that the Rashba
SO interaction is independent of the crystallographic direction, and is
invariant under rotation about the normal axis of the 2DEG plane. The spatial
behavior of the injected spin precessing with space (rather than time) and
rotating about the perpendicular field [red bold arrows in Fig. 1(a), and also
in (b) and (c)] is clearly observed. Since the Rashba field is always normal
to the electron wave vector, the injected spinexhibits an upright precession
when the electron propagates parallel to the polarization of the injected
spin, while the precession is suppressed when propagating perpendicular to the
injected spin.%
\begin{figure}
[ptb]
\begin{center}
\includegraphics[
trim=1.598800in 0.000000in 1.599576in 0.000000in,
height=6.9263in,
width=2.8893in
]%
{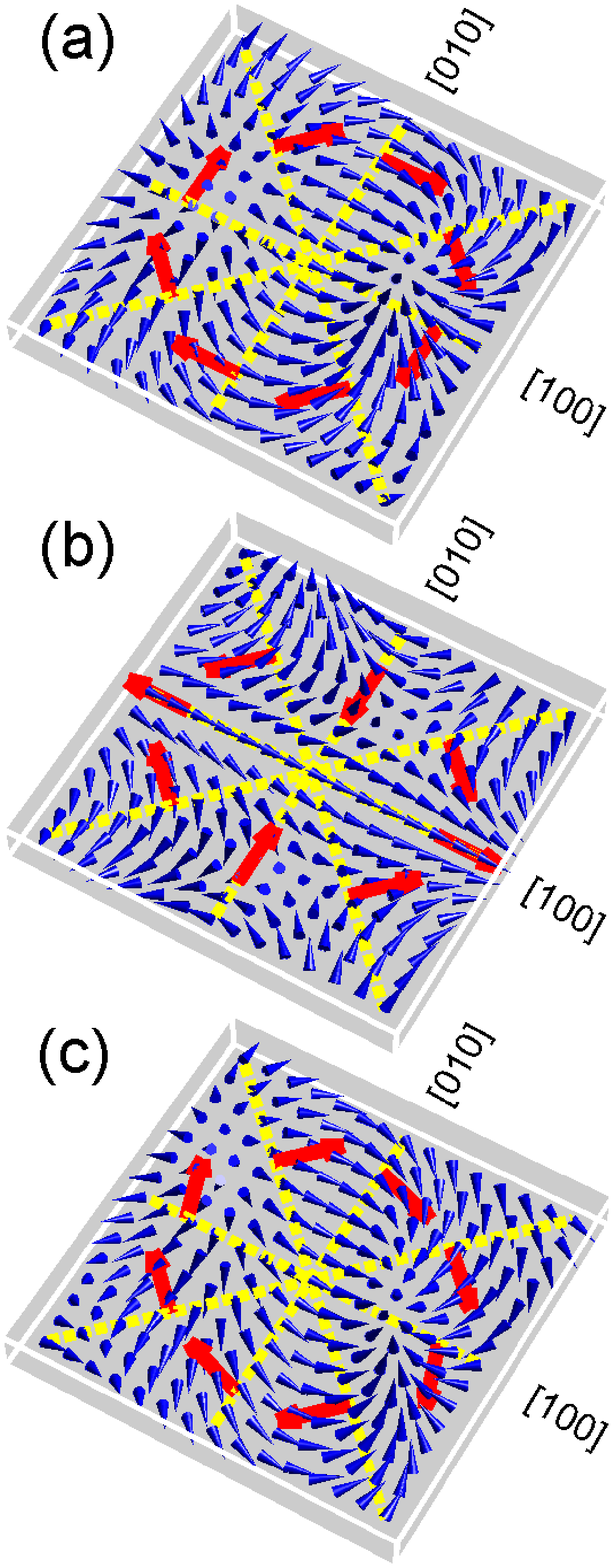}%
\caption{Spin precession due to (a) SIA, (b) BIA, and (c) both SIA and BIA for
[001]-grown 2DEG planes. The spin injection point is set on the center of the
2DEG. Red bold arrows indicate the directions of the effective magnetic
fields. In all cases, the length dimensions are in arbitrary units. The ratio
of the Rashba and Dresselhaus coupling strengths is set 3/1 in (c).}%
\end{center}
\end{figure}

In the pure Dresselhaus case [Fig. 1(b)], the injected spin encounters another
type of effective magnetic field, and hence the spatial behavior is quite
different from the Rashba one. The generated field directions are no more
perpendicular to the wave vector, except those along $\pm$[1$\pm$10], on which
the Dresselhaus spin precession exhibits the same spatial behavior with the
Rashba one. In addition, the rotational invariance about z, contrary to the
Rashba case, is broken, and the crystallographic direction dependence hence
comes in.

Indeed, in the composite case shown in Fig. 1(c), where we set the coupling
ratio $\alpha/\beta=3$, the total effective magnetic field remains
perpendicular to the electron propagation along $\pm$[1$\pm$10]. In the design
of the Datta-Das spin-FET, one can see that if the channel direction is
arbitrarily chosen, e.g., along [100], the spin precession may be slanted,
thus lowering the efficiency of the tunability of the resulting current via
the gate voltage.

\section{Datta-Das spin-FET}

To be more specific on the influence by the BIA on the Datta-Das spin-FET, let
us now consider an ideal one-dimensional channel confined in a zinc-blende
type semiconductor, where both the SIA and BIA are present. Setting the
Dresselhaus SO strength $0.9\times10^{-11}$ $%
\operatorname{eV}%
\operatorname{m}%
$, we analyze the projection of the spin vector on the drain magnetization
$\mathbf{m}$ (set parallel with the channel direction as the design of the
Datta-Das spin-FET) for respectively the $\alpha$-dependence and position
dependence, along [100] and [1$\pm$10].

In Fig. 2(a), the spin detection point is fixed on the end of the channel with
the Rashba SO strength tuned from 0 to $3\times10^{-11}$ $%
\operatorname{eV}%
\operatorname{m}%
$, which corresponds to typical values for InGaAs 2DEGs \cite{10}. The channel
length is set $0.5$ $%
\operatorname{\mu m}%
$ and the effective mass is $0.03$ times the electron rest mass \cite{10}.
Clearly, only the [1$\pm$10] curves exhibit maximized oscillation due to the
upright spin precession. When choosing the channel direction as [100], the
electron encounters tilted effective magnetic field, and the oscillation
amplitude decreases, until the SIA strength is much stronger than the BIA.
This may lower the spin signal of the Datta-Das transistor quite severely and
hence only [1$\pm$10] are the proper candidates for the channel direction of
the Datta-Das spin-FET.%
\begin{figure}
[ptb]
\begin{center}
\includegraphics[
height=1.4364in,
width=3.3762in
]%
{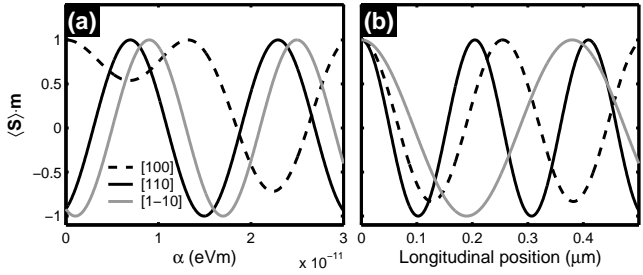}%
\caption{Projection of spin vector on the drain magnetization, set parallel to
the channel direction, for (a) $\alpha$ dependence and (b) position dependence
examination.}%
\end{center}
\end{figure}

Due to the anisotropy of the spin-splitting strength [Eq. (7)], the precession
period, in fact, varies with the channel direction. For $\alpha$ and $\beta$
of the same sign, the spin-splitting along [110] is strongest: $\gamma\left(
\pi/4\right)  =\left\vert \alpha+\beta\right\vert $ while that along [1-10] is
weakest: $\gamma\left(  -\pi/4\right)  =\left\vert \alpha-\beta\right\vert $.
Fixing the Rashba strength $\alpha$ as $3\times10^{-11}$ $%
\operatorname{eV}%
\operatorname{m}%
$, we analyze the spin projection on the longitudinal position varied from 0
to $0.5$ $%
\operatorname{\mu m}%
$. Again, the amplitude of the [100] curve does not reach the maximum since
the precession is slanted, while the [1$\pm$10] curves do. However, the
precession period for the [1-10] curve is almost twice the [110] curve since
the SO coupling ratio along these two directions is $\left\vert \alpha
+\beta\right\vert /\left\vert \alpha-\beta\right\vert =1.86$.

\section{Conclusion}

In conclusion, we have derived the analytical expression for the spin
expectation of a point-injected electron, using a standard quantum mechanical
approach. The obtained formulae are useful for analyzing the spatial behaviors
of the spin precession caused by individually the SIA and the BIA, and even
the spin vectors in the whole 2DEG channel, considering both the SIA and BIA
mechanism. The dependence of the detection position, SO coupling strengths,
and orientation angle of the injected spin-polarization are included in the
formulae, and hence provides a wide variety to investigate the feasibility of
the Datta-Das transistor for more details.

In the analysis of the spin precession behaviors, the Rashba case shows
rotation symmetry about the normal axis of the 2DEG plane, while this symmetry
is broken in the Dresselhaus case. Furthermore, we found that the injected
spin always encounters a perpendicular effective magnetic field along $\pm
$[1$\pm$10] in the composite case, and these two directions are therefore
candidates for the Datta-Das spin-FET channel directions.

By considering an ideal one-dimensional channel with injected spin polarized
parallel to the channel direction, we also re-examined the influence due to
the BIA on the Datta-Das spin-FET. We demonstrated the significance of proper
choice of the channel direction. Choosing directions other than $\pm$[1$\pm
$10] may severely decrease the spin signal collected by the drain contact. In
addition, electrons transiting along [110] are found to encounter stronger SO
coupling than along [1-10] in the case of $\alpha\beta>0$. Thus we conclude
that [110] may be the best candidate for the channel direction of the
Datta-Das spin-FET.

\begin{acknowledgments}
This work was supported by the Republic of China National Science Council
Grant No. 94-2112-M-002-004.
\end{acknowledgments}

\end{document}